# Enhanced Sensitivity Beam Emission Spectroscopy System for Nonlinear Turbulence Measurements[1]

Deepak K Gupta[2], Raymond J Fonck, George R McKee,

David J Schlossberg and Morgan W Shafer

*University of Wisconsin--Madison, 1500 Engineering Drive, Madison, Wisconsin-53706, USA*

## Abstract

An upgraded Beam Emission Spectroscopy (BES) system has been deployed to access low amplitude turbulence regions near internal transport barriers on the DIII-D tokamak. Sixteen high sensitivity channels are being installed. A significant increase in total signal to noise is achieved by: 1.) Increased spatial volume sampling tailored to known turbulence characteristics; 2.) An increased throughput spectrometer assembly to isolate the local beam fluorescence, coupled to new large-area photoconductive photodiodes; 3.) A new sharp edge interference filter designed to optimize detection of the beam emission plus a significant fraction of the thermal deuterium charge exchange. A new data acquisition system has been installed, providing an 8 times increase in integration time or an increased sample rate. Preliminary results from the upgraded system show a signal enhancement of greater than an order of magnitude. A clear broadband density fluctuation signal is observed in H-mode discharges with the upgraded BES system, demonstrating the significant performance enhancement.

---

[2] Email: dpk011@yahoo.com or dpk011@gmail.com

## Introduction

Beam Emission Spectroscopy (BES)[1] has been extensively utilized to study turbulence and associated transport in magnetically confined high temperature plasmas. It has been used to study zonal flows[2], impurity-induced turbulence suppression in radiative improved (RI) mode[3], nonlinear energy transfers[4] and other turbulence characteristics. One limitation of the original BES system[5] on the DIII-D tokamak has been limited signal to noise, inhibiting studies of low-turbulence regions such as the core of high-performance (advanced) tokamak discharges. For sufficiently low amplifier noise, sensitivity of the BES system is determined by the statistical noise in detected photon flux.[6] The square root dependence of this statistical noise on detected photon flux require many fold enhancement in photon flux to achieve a significant increase in Signal-to-Noise Ratio (SNR). To improve performance and expand turbulence studies to nonlinear processes, a significant upgrade of the BES density fluctuation diagnostic system on DIII-D tokamak has been initiated. Sensitivity is expected to increase by a factor of 5-10, extending the studies deeper into the core of tokamak plasmas and facilitate new ones like direct measurement of turbulence driven fluxes, studies of internal transport barriers and 2D imaging for visualization of core turbulence. This paper presents a summary of various hardware upgrades performed to achieve this higher sensitivity. Preliminary results on turbulence measurements in H-mode discharges from upgraded system are presented and compared with similar measurements from the original system.

## BES Hardware Upgrade

BES is a crossbeam, volume-sampling density fluctuation diagnostic based on the measurement of deuterium neutral beam fluorescence ($D_\alpha$, n=3-2 transition near $\lambda_o$=656 nm) induced by collisions with background plasma ions and electrons. Details of the original 32 spatial channels BES system installed at DIII-D tokamak are described in reference 5. The sensitivity (or signal-to-noise ratio, SNR) of the next generation BES upgrade is improved by: i) increased spatial plasma volume sampling tailored to known turbulence characteristics, ii) increased throughput spectrometer assembly coupled to new large area photoconductive PIN-photodiodes, and iii) high-transmission, sharp-edge interference filters designed to integrate a significant fraction of $D_\alpha$ thermal charge exchange spectra, in addition to the beam emission.

An optics assembly viewing a heating neutral beam on DIII-D tokamak images beam emission light onto a set of fiber optics that convey the light 40 meters to a remotely-located spectroscopy laboratory. A plasma region is focused using a f/2, 40 cm focal length objective lens onto fiber bundles, mounted on a remotely-controlled radially scanable fiber mount. The intersection of the optical sightline with the neutral beam volume is aligned to a magnetic flux surface to achieve good spatial resolution in the radial and poloidal planes ($\Delta r \approx \Delta Z \approx 1$ cm). Each channel in the original optical fiber bundles consists of four-1mm diameter fibers, while the upgraded system consists of new fiber bundles with eleven-1mm diameter fibers arranged in a 4:3:4 pattern, as shown in Fig.1(a). This new fiber bundle design is more optimally aligned to the radial and poloidal asymmetry of the turbulent eddies.[7] These changes improve the signal collection by 2.75 times per channel compared to original fiber bundle, and the areal coverage in the image plane is also increased from 19% to 79%. Likewise the channel-to-channel separation is nearly the same (radial separation is reduced from about 1.1 cm to 0.9 cm, and poloidal separation is increased from 1.1 to 1.2 cm). The wavenumber sensitivity is calculated using the full 3D viewing geometry, neutral beam profile and magnetic field geometry. There is a slight reduction in the wavenumber sensitivity, as shown in Fig.1(b), however this reduction is modest given the substantial increase in photon flux. Optical axis folding mirrors (located between the objective lens and the fibers to avoid a toroidal field coil) have been replaced with enhanced and protected silver-coated mirrors having 98-99% reflectivity across the visible spectrum. Finally, an anti-reflective (AR) coating on each end of the optical fibers should further improve transmission by about 8%.

The high throughput spectrometers consist of appropriate lens assembly, interference filters and low noise detectors to isolate the local beam fluorescence from the collected light signal.[5] The collection efficiency of the spectrometers is improved by using 50mm diameter, f/1.5 plano-convex collimating lenses to collect the light from the wider fiber bundles. The collimated beam is normally incident on new 5cm diameter interference filters, which are designed to also transmit a fraction of the thermal deuterium charge exchange lines on the blue side of edge $D_\alpha$ emission (as shown in Fig.2). These large filters have increased absolute peak transmission (nearly 80%), and a sharper spectral edge compared to the ones used in original BES system. This optimizes transmission of the charge exchange emission and suppresses the intense but unwanted edge recycling light (transmission at $D_\alpha$ is < 1%). The filters are designed to be use at near normal incidence to reduce the spectral blurring that results from angle tuning. The optimized filter is designed to provide at least a 1.5 times increase in the signal strength relative to filters in the original system. The collimated, filtered light beam is focused on the detectors using an ultra fast f/0.58 AR-coated aspheric lens. The fast aspheric lens facilitates use of small photo detector and thus helps in keeping the amplifier noise low.

In the original BES system, 0.85 $mm^2$ square photoconductive PIN Photodiodes (PPD) were used. To minimize the amplifier noise, the photodiode and first-stage preamplifier FET are cryogenically cooled.[5,6] The voltage noise (i.e., e-noise), a dominant source of noise in preamplifier circuits[6], is associated with the detector capacitance and thus detector area. Hence to keep the amplifier noise minimum, use of the smallest detector that accommodates all the available light is necessary. For upgraded BES system, an improvement in SNR can be estimated for the range of power ratio of photon noise to amplifier noise, ρ, of original BES system by using the analytical relation

$$\frac{SNR_U}{SNR_O} = \frac{g_\gamma}{F}\left[\frac{1+\rho}{(D^2/M^2Fg_\gamma)+\rho}\right].$$

Where $SNR_U$ and $SNR_O$ are the Signal-to-Noise power ratio of upgraded and original BES system respectively; $g_\gamma$ is the photon flux gain from the improved optics and $D$ is the amplifier noise gain (e.g., due to increase in photodiode area). For Avalanche Photodiode (APD), the intrinsic gain, $M$ and the excess noise factor, $F$, have values greater than one; however, for PPD values of $M$ and $F$ are unity. The APD offers a high intrinsic gain, which potentially eliminates the need for cryogenically cooled preamplifiers and associated e-noise. However, the excess noise factor in an APD effectively offsets this advantage for the plasma turbulence and BES signal parameters of interest. The high quantum efficiency (≈85%) PPD along with the increased photon flux offer an advantage up to a factor of 3 or more over APD for the BES upgrade and are thus being utilized. For upgraded BES system, a 2.91mm$^2$ rectangular PPD collects the full image of the new fiber bundle when rotationally aligned. The input capacitance of the high quality FET in first-stage preamplifier is the dominating capacitance in defining the total capacitance contribution toward the e-noise. Hence even after about three times increase in detector area and thus its capacitance (which adds to the FET capacitance for the total capacitance towards e-noise), only a modest 20-50% increase in the amplifier noise is observed with respect to the e-noise in original BES system.

A 14-bit resolution Linux-based data acquisition system has replaced the original 12 bit CAMAC digitizers.[8] The 32 channel high-speed digitizers (from D-tAcq Solutions) can acquire up to 4 seconds of data at 1 MHz sampling rate, compared with 0.5 seconds for original CAMAC digitizers. This effectively improves SNR by about 3 times for ensemble-averaged measurements of stationary turbulence and also reduces the need for shot repetition.

## Preliminary Results

The sixteen high-sensitivity spatial channels have been deployed at DIII-D tokamak, which measure a 4x4 grid in the radial-poloidal plane (see Fig 1(a)). Data has been acquired simultaneously with the upgraded channels and the original BES channels for a H-mode discharge. Signal strengths from the upgraded BES system are more than 10 times higher compared to those acquired with the old BES system, as shown in Fig. 3., thus achieving and surpassing the predicted signal enhancement. Additional factors such as improved-transmission fiber optics, improved coupling of fiber image to detector, long-term degradation of the interference filter transmission, coated optics or detectors, etc also contributed to this observed signal improvement. Based on the observed increase in total signal, it is estimated using the analytical formula described in last section that the SNR for fluctuation power measurements will improve in the range of 14 to 30 depending upon the plasma properties and frequency dependence of the fluctuation and noise power spectra.

To demonstrate the capabilities of the substantially increased signal levels, the coherency spectra of raw signals from two adjacent channels from the upgraded and original BES systems are compared in Figure 4. Measurements from the upgraded system show a clear broadband density fluctuation signal in the 60-200kHz range that is simply not observable with the original system in H-mode discharges. The signal in the 0-50kHz range appears to result from ELMs, edge-turbulence imprinted on the beam or other source of neutral beam modulation rather than from local density fluctuations. Common-mode analysis techniques may be used to isolate and subtract these effects.[9]

The substantial increase in measured light signal levels, and corresponding quantum leap in signal-to-noise ratio, as well as the demonstrated measurement of a strong fluctuation signal in H-mode discharges demonstrates significant new capability with the upgraded BES density fluctuation diagnostic system. Longer-term plans are to deploy up to 64 high-performance channels for wide area, high-sensitivity turbulence imaging capability in the core of advanced tokamak discharges along with detailed studies of the nonlinear characteristics of turbulence driving cross-field plasma transport.

This work is supported by U.S. DoE grant no. DE-FG03-96ER54373

**Figures:**

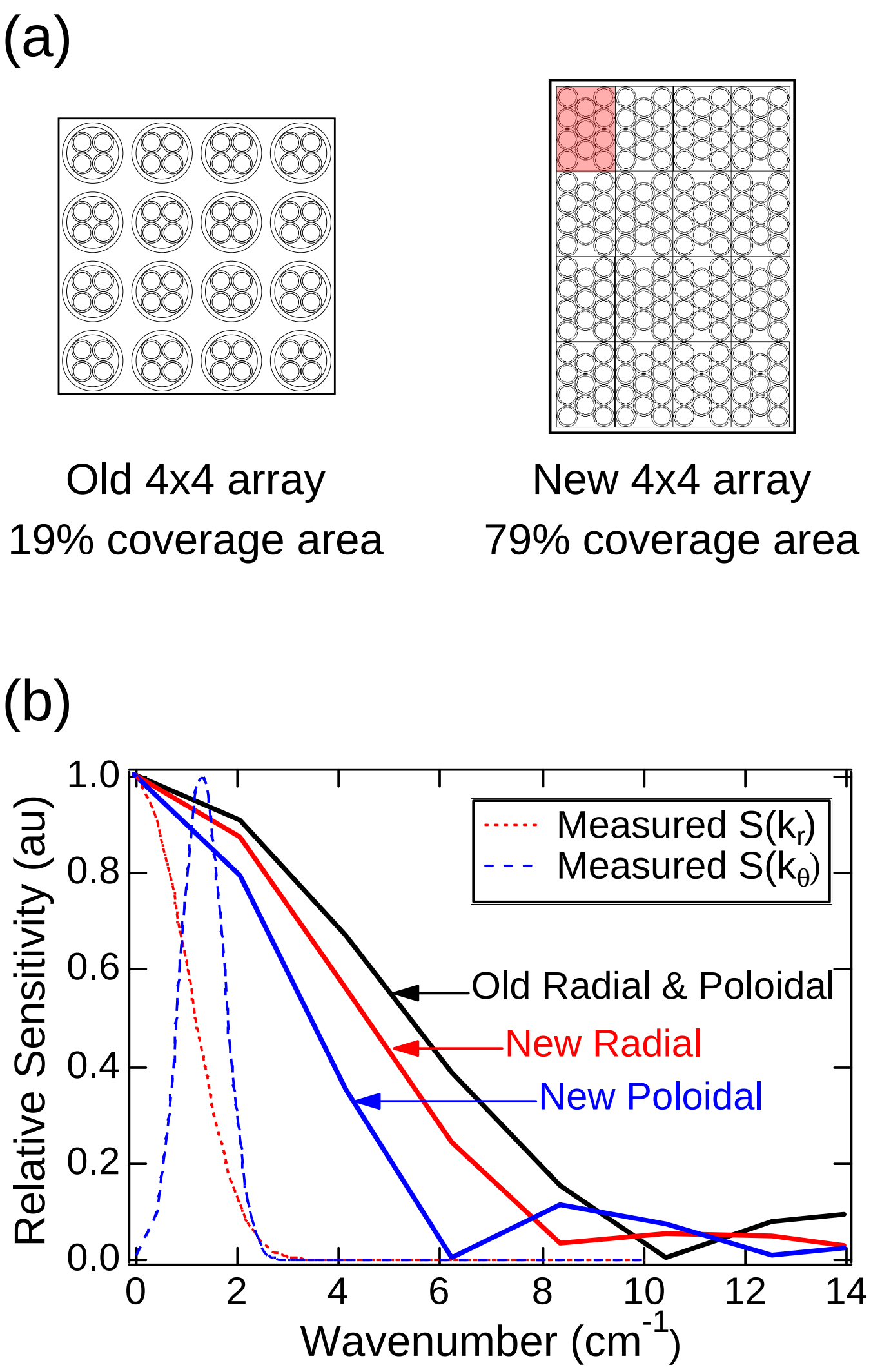

**Figure 1: (a) Comparison of original fiber bundles (4 fibers, square arrangement) and new fiber bundles (11 fibers, close packed arrangement) in 4x4 array. To optimally align natural turbulence eddies, radial (channel to channel) spacing is reduced by 2mm and poloidal spacing is increased by 1mm. (b) Comparison of wavenumber sensitivity for new and old sample volume shows that upgraded system has sufficient resolution for typical measured turbulence data.**

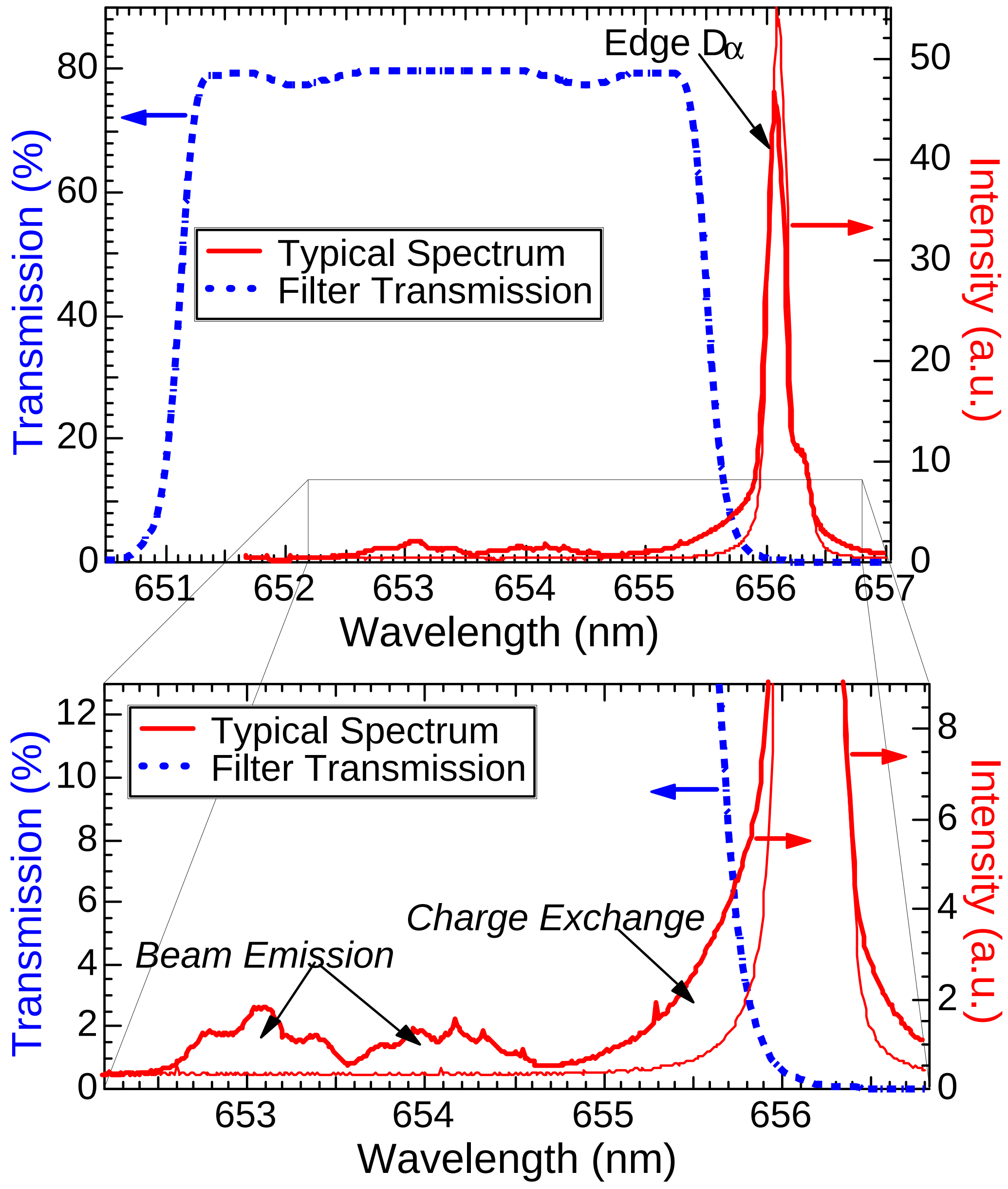

**Figure 2: A new interference filter transmission spectrum includes Doppler-shifted beam emission as well as a significant fraction of deuterium thermal charge exchange. The filter eliminates much of the edge recycling $D_\alpha$ emission.**

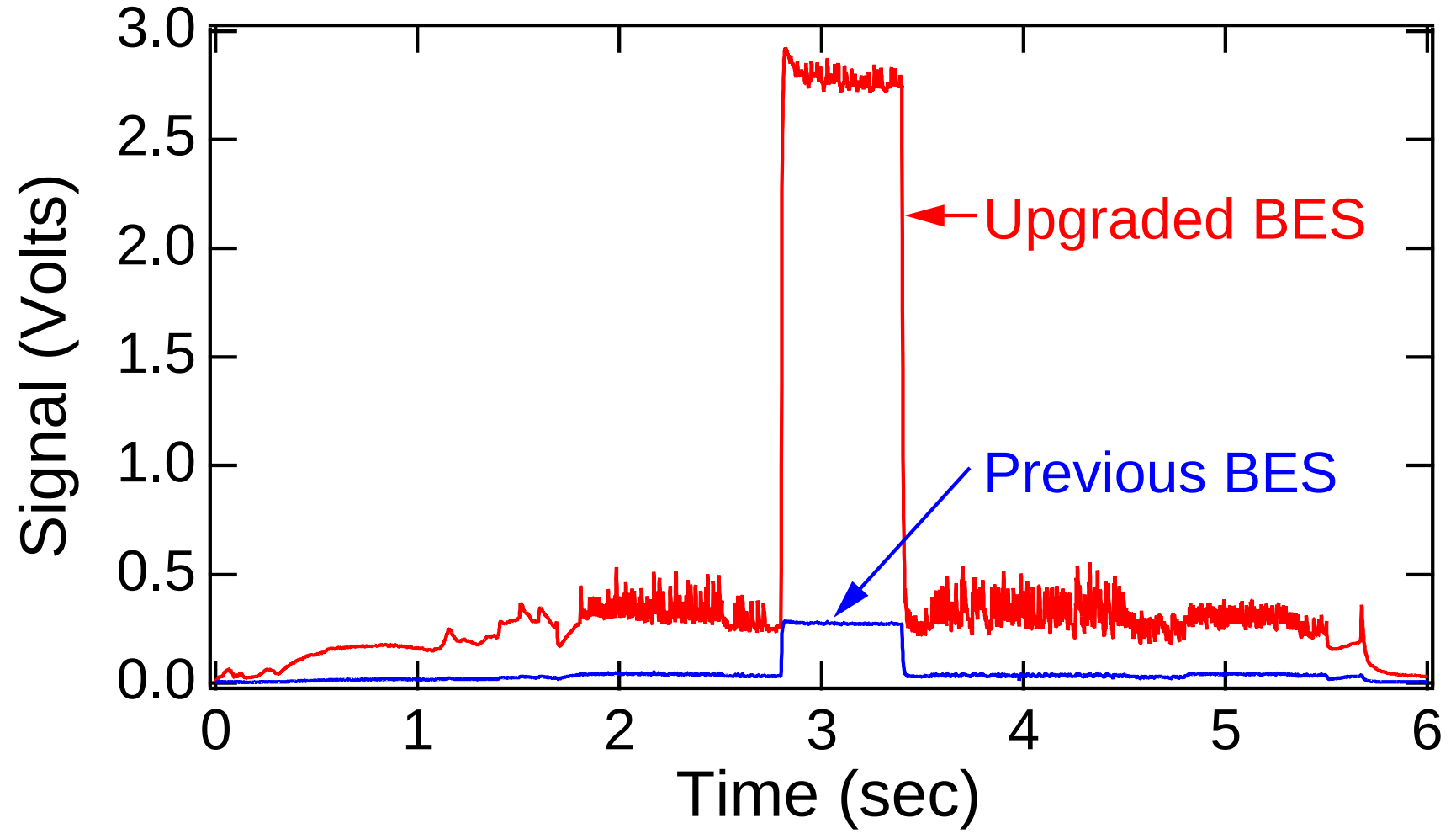

**Figure 3: A comparison of signal level of original and upgraded BES system shows an order of magnitude increase in observed signal. Neutral beam is on from 2.8 seconds to 3.4 seconds.**

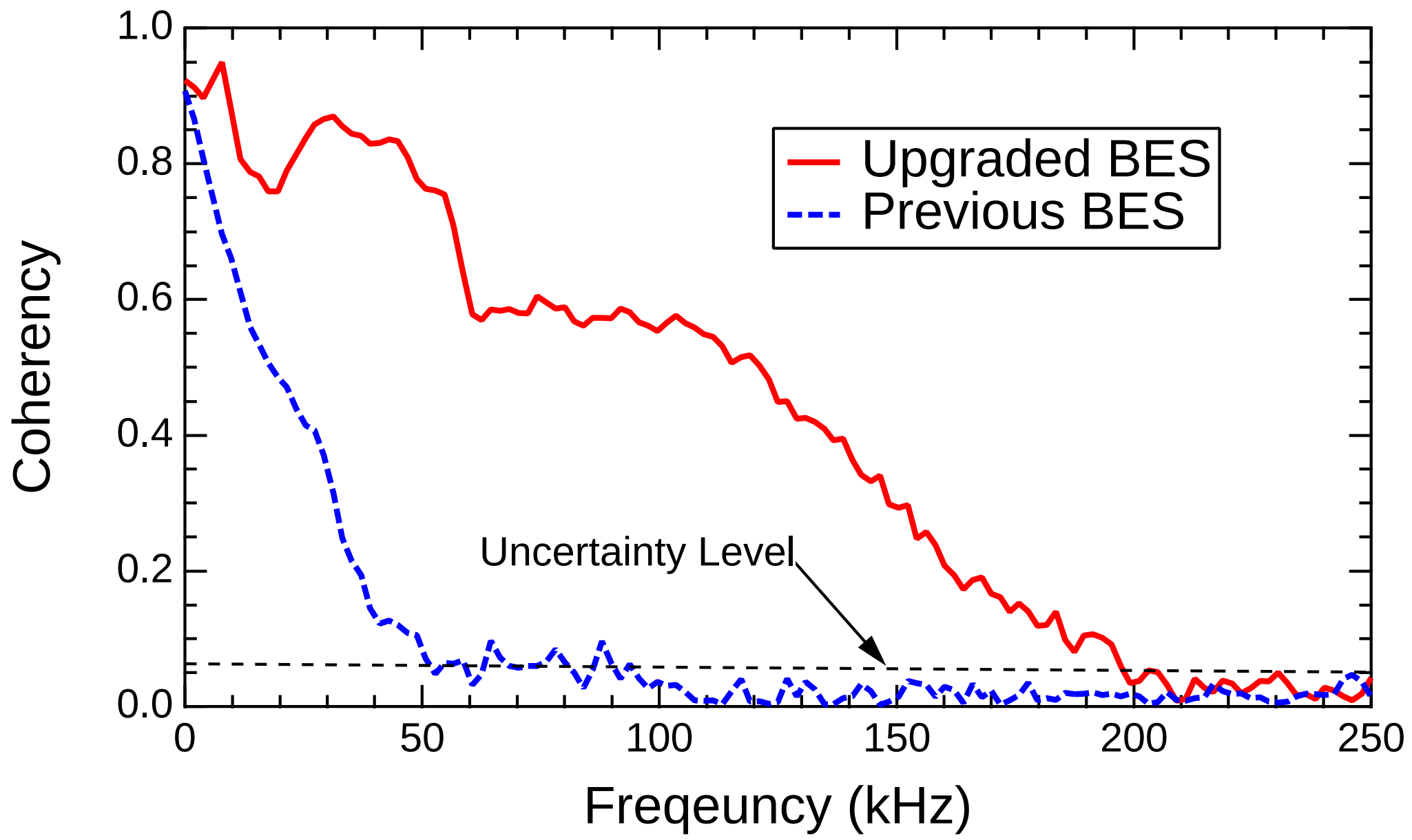

**Figure 4: Comparison of density fluctuation coherency spectra of raw signals (Δx≈1 cm) from the upgraded and original BES channels, acquired near r/a=0.7 in a H-mode discharge.**